\documentclass[10pt]{article}
\usepackage{graphicx}
\usepackage{subfigure}
\usepackage{lineno}

\def\Title#1{\begin{center} {\Large #1 } \end{center}}
\def\Author#1{\begin{center}{ \sc #1} \end{center}}
\def\Address#1{\begin{center}{ \it #1} \end{center}}

\def\met{\ensuremath{E_{\mathrm{T}}^{\mathrm{miss}}}}
\def\pt{\ensuremath{p_{\mathrm{T}}}} % Subscript roman not italic (EE)                                                                                                              

\newcommand\pubblock{\rightline{\begin{tabular}{l} Proceedings of the Second Annual LHCP\\ \pubnumber\\
         \pubdate  \end{tabular}}}

\newenvironment{Abstract}{\begin{quotation} \begin{center} 
             \large ABSTRACT \end{center}\bigskip 
      \begin{center}\begin{large}}{\end{large}\end{center} \end{quotation}}

\newenvironment{Presented}{\begin{quotation} \begin{center} 
             PRESENTED AT\end{center}\bigskip 
      \begin{center}\begin{large}}{\end{large}\end{center} \end{quotation}}

\def\Acknowledgements{\bigskip  \bigskip \begin{center} \begin{large}
             \bf ACKNOWLEDGEMENTS \end{large}\end{center}}
\def\beq{\begin{equation}}
\def\eeq#1{\label{#1}\end{equation}}
\def\eeqn{\end{equation}}

\def\beqa{\begin{eqnarray}}
\def\eeqa#1{\label{#1}\end{eqnarray}}
\def\eeqan{\end{eqnarray}}

\let\bar=\overbar

\def\Dslash{\not{\hbox{\kern-4pt $D$}}}
\def\dslash{\not{\hbox{\kern-2pt $\del$}}}

\def\msb{{\bar{\ssstyle M \kern -1pt S}}}

\usepackage{color}

\newcommand\pubnumber{ ATL-PHYS-PROC-2014-106 }
\newcommand\pubdate{\today}

\def\affiliation{
On behalf of the ATLAS and CMS Collaborations, \\
ARC Centre of Excellence for Particle Physics at the Terascale\\
School of Physics\\
The University of Melbourne, Victoria 3010, Australia}

\begin{document}
%\linenumbers

% large size for the first page
\large
\begin{titlepage}
\pubblock

%% Change the title, name, abstract
%% Title 
\vfill
\Title{ Overview of Triple and Quartic Gauge Coupling Measurements at the LHC  }
\vfill

%  if you need to add the support use this, fill the \support definition above. 
%   \Author{ FIRSTNAME LASTNAME \support }
\Author{ Takashi Kubota }
\Address{\affiliation}
\vfill
\begin{Abstract}

Scrutiny of the structure of electroweak gauge boson self-interactions through triple and quartic gauge boson couplings (TGCs and QGCs) constitutes an important part of the physics program of the Large Hadron Collider (LHC).
Triboson production and vector boson scattering (VBS) are directly sensitive to QGCs while vector boson fusion (VBF) offers a new window in the study of TGCs, which is complementary to conventional measurements using diboson production.
In this contribution, an overview of recent TGC and QCG measurements using triboson production, VBS and VBF by the ATLAS and CMS experiments at the LHC is presented.

\end{Abstract}
\vfill

% DO NOT CHANGE 
\begin{Presented}
The Second Annual Conference\\
 on Large Hadron Collider Physics \\
Columbia University, New York, U.S.A \\ 
June 2-7, 2014
\end{Presented}
\vfill
\end{titlepage}
\def\thefootnote{\fnsymbol{footnote}}
\setcounter{footnote}{0}
%

% normal size for the rest
\normalsize 

%% Your paper should be entered below. 

\section{Introduction}

The non-abelian nature of the electroweak sector predicts the self-interaction structure of electroweak gauge bosons in the form of triple and quartic gauge boson couplings (TGCs and QGCs).
While the study of these couplings offers an important test of the Standard Model of particle physics (SM), QGCs are additionally connected to electroweak symmetry breaking sector, 
together with the Higgs boson, ensuring unitarity at high energies for scattering processes.

The proton-proton collision data provided by the Large Hadron Collider (LHC), Geneva, offers excellent sensitivity in hunting for deviations from the SM in QGCs through triboson production and vector boson scattering (VBS) while providing another window for the study of TGCs through vector boson fusion (VBF).
%, which is complementary to conventional measurements using diboson productions.
In this paper, an overview of TGC and QGC measurements using triboson production, VBS and VBF by the ATLAS and CMS experiments~\cite{ATLAS,CMS} is presented.

\section{TGC and QGC Measurements at the LHC}

The following TGC and QGC measurements performed by the ATLAS and CMS experiments are presented.
\begin{itemize}
{\setlength\itemindent{25pt} \item[$\bullet$~{\bf \ref{ATLAS_ssWWjj}}] A QGC measurement using same sign $WW$ production with two jets~\cite{ATLAS_ssWWjj};}
{\setlength\itemindent{25pt} \item[$\bullet$~{\bf \ref{CMS_ggWW}}] A QGC measurement using the exclusive $\gamma\gamma \rightarrow WW$ production~\cite{CMS_ggWW};}
{\setlength\itemindent{25pt} \item[$\bullet$~{\bf \ref{CMS_VVV}}] A QGC measurement using the $WW\gamma/WZ\gamma$ production~\cite{CMS_VVV};}
{\setlength\itemindent{25pt} \item[$\bullet$~{\bf \ref{EWZjets}}] A TGC measurements using the VBF $Z$ boson production with two jets~\cite{ATLAS_EWZjets,CMS_EWZjets}.}
\end{itemize}

\subsection{Same-sign $WW$ production with two jets}\label{ATLAS_ssWWjj}

The scattering of two $W$ bosons is a key process in testing the nature of electroweak symmetry breaking.
Even after the discovery of an SM-like Higgs boson by the ATLAS and CMS experiments~\cite{ATLAS_Higgs,CMS_Higgs} many physics scenarios predict enhancements in VBS~\cite{ATLAS_ssWWjj_par1,ATLAS_ssWWjj_par2}.
At the LHC, the process can be idealized as an interaction of $W$ bosons radiated from initial state quarks yielding a final state with two $W$ bosons and two jets ($WWjj$).
$WWjj$ processes that involve exclusively weak interactions at Born Level is referred to as electroweak production, which includes VBS contributions, while those involving the strong interaction are referred to as strong production.
In the case of same-electric charge $WW$ production, the strong production does not dominate the electroweak production, making this channel an ideal choice for initial studies on VBS.

An analysis is performed by the ATLAS experiment using $20.3~{\rm fb}^{-1}$ of data at $\sqrt{s}$ = 8 TeV~\cite{ATLAS_ssWWjj}.
In the analysis, a fiducial region for electroweak production is defined as follows:
exactly two prompt charged leptons are required with the same electric charge,
transverse momentum $\pt > $ 25 GeV, $|\eta| < 2.5$, invariant mass $m_{\ell \ell} > $ 20 GeV and angular separation $\Delta R_{\ell\ell} > $ 0.3;
at least two jets reconstructed with the anti-$k_{t}$ algorithm with a distance parameter R = 0.4 with $\pt > $ 30 GeV, $|\eta| < $ 4.5, and separated from the leptons by $\Delta R_{\ell j} > $ 0.3;
the invariant mass of two jets with the largest $\pt~(m_{jj})$ must be larger than 500~GeV;
the magnitude of the missing transverse momentum ($\met$) reconstructed in the final state must be greater than 40 GeV;
%To reduce the dependence on QED radiation, lepton momenta include contributions from photons within $\Delta R $ = 0.1 of the lepton direction.
the separation in rapidity of the two jets with largest $\pt$ is required as $|\Delta y_{jj}| > $ 2.4.

\begin{figure}[htp]
  \begin{center}
    \subfigure[]{\includegraphics[width = 0.48\textwidth]{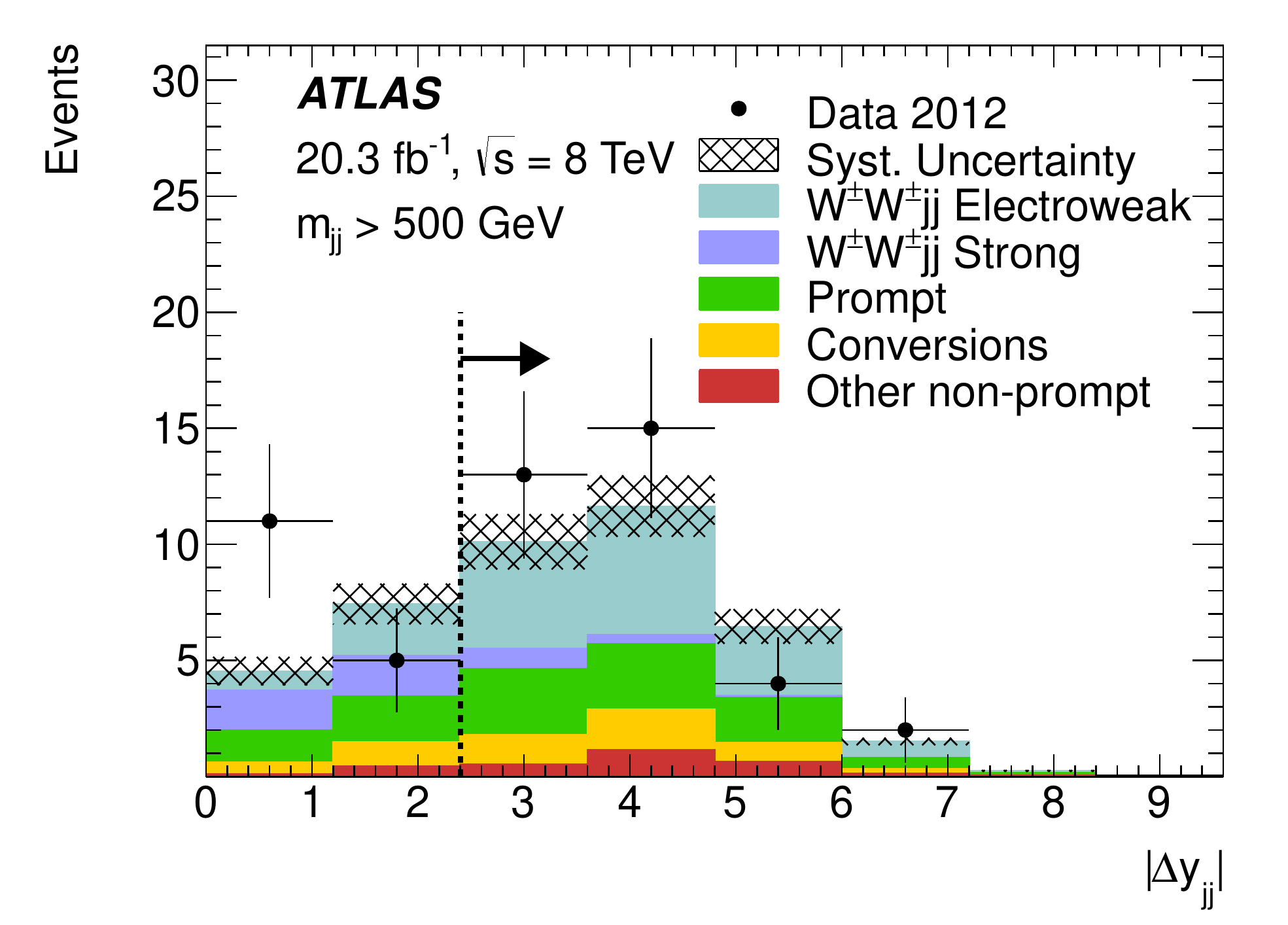}}
    \subfigure[]{\includegraphics[width = 0.48\textwidth]{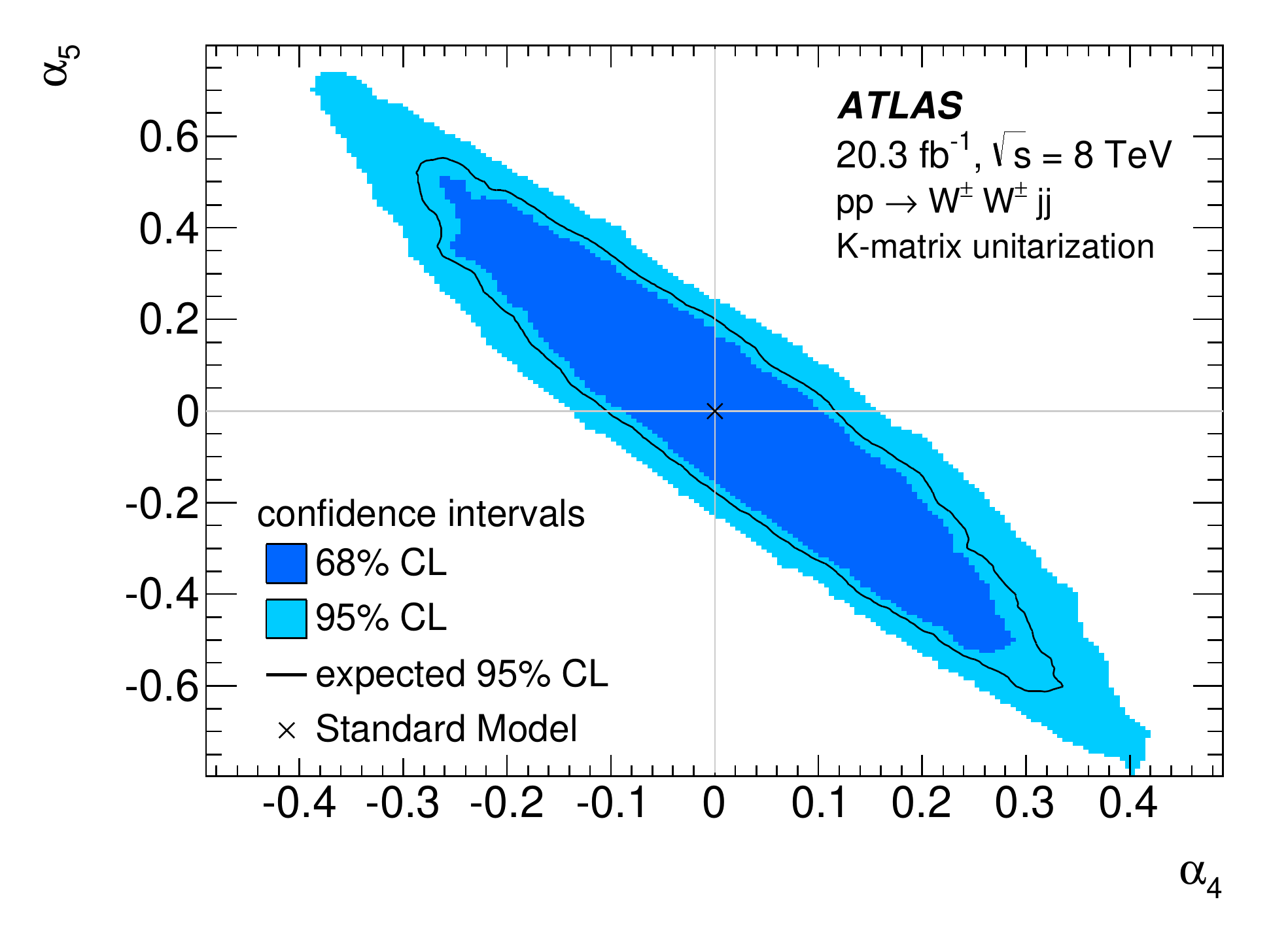}}
    \caption{(a) The $|\Delta y_{jj}|$ distribution after all the fiducial region cuts other than the cut on $|\Delta y_{jj}|$ and (b) the limits on aQGC parameters ($\alpha_{4}$, $\alpha_{5}$) set in the same-sign $WWjj$ analysis~\cite{ATLAS_ssWWjj}.}
    \label{fig:wwjj}
  \end{center}
\end{figure}

Figure~\ref{fig:wwjj} (a) shows the $|\Delta y_{jj}|$ distribution after the fiducial region cuts other than the $|\Delta y_{jj}| > $ 2.4 requirement.
The measured fiducial cross-section for electroweak $WWjj$ production, including interference with strong production is $\sigma^{\rm fid} = 1.3 \pm 0.4 ({\rm stat}) \pm 0.2 ({\rm syst})$ fb.
This measured fiducial cross-section is used to set limits on anomalous QGCs (aQGCs) affecting vertices with four interacting $W$ bosons.
The {\sc WHIZARD} event generator is used to generate $WWjj$ events with aQGCs using the K-matrix unitarisation method,
and deviations from the SM, which includes a SM Higgs with $m_{\rm H}$ = 126 GeV, are parameterised in terms of two parameters ($\alpha_{4}, \alpha_{5}$)~\cite{ATLAS_ssWWjj_kmatrix}.
The expected and observed 95\% confidence level (C.L.) intervals derived from the profile likelihood function are shown in Figure~\ref{fig:wwjj} (b).
The one-dimensional projection at $\alpha_{5,4}$ = 0 is respectively $-0.14 < \alpha_{4} < 0.16$ and $-0.23 < \alpha_{5} < 0.24$ compared to an expected $-0.10 < \alpha_{4} < 0.12$ and $-0.18 < \alpha_{5} < 0.20$. 

\subsection{Exclusive $\gamma\gamma \rightarrow WW$ production}\label{CMS_ggWW}

The detection of high-energy photon interactions at the LHC opens up the possibility of interesting and novel research.
In particular, measurements of the two-photon production of a pair of $W$ bosons provide sensitivity to aQGCs.
An analysis using $5.05~{\rm fb}^{-1}$ of data at $\sqrt{s}$ = 7 TeV taken by the CMS detector is performed~\cite{CMS_ggWW}.
The $\mu^{\pm}e^{\mp}$ final state is used to search for fully exclusive $pp \rightarrow p^{(\ast)}W^{+}W^{-}p^{(\ast)}$ production, where $p^{(\ast)}$ denotes either of very forward-scattered intact protons or a proton dissociated into a low-mass systems that escape detection.
As in either case protons escape detection, this process is characterized by a primary vertex formed from a $\mu^{\pm}e^{\mp}$ pair with no other tracks, with large transverse momentum ($p_{{\rm T},\mu^{\pm}e^{\mp}}$) and large invariant mass ($m_{\mu^{\pm}e^{\mp}}$).
%The two-photon signal $\gamma\gamma \rightarrow W^{+}W^{-}$ is therefore comprised of both the elastic and inelastic contributions.

\begin{figure}[htp]
  \begin{center}
    %\subfigure[]{\includegraphics[width = 0.96\textwidth]{figures/ggww/EventDisplays.png}}
    \subfigure[]{\includegraphics[trim=9.5cm 0.8cm 0cm 9.5cm, width = 0.40\textwidth]{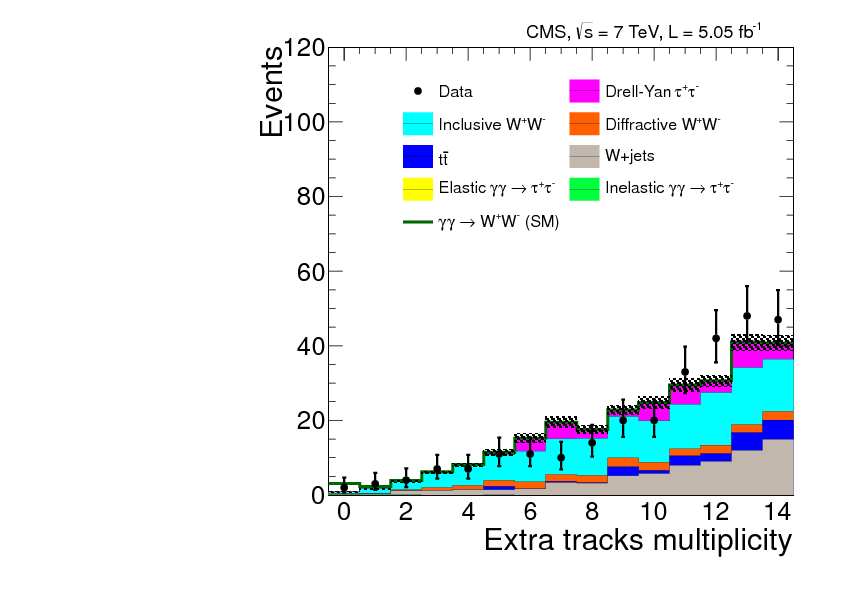}}
    \subfigure[]{\includegraphics[width = 0.40\textwidth]{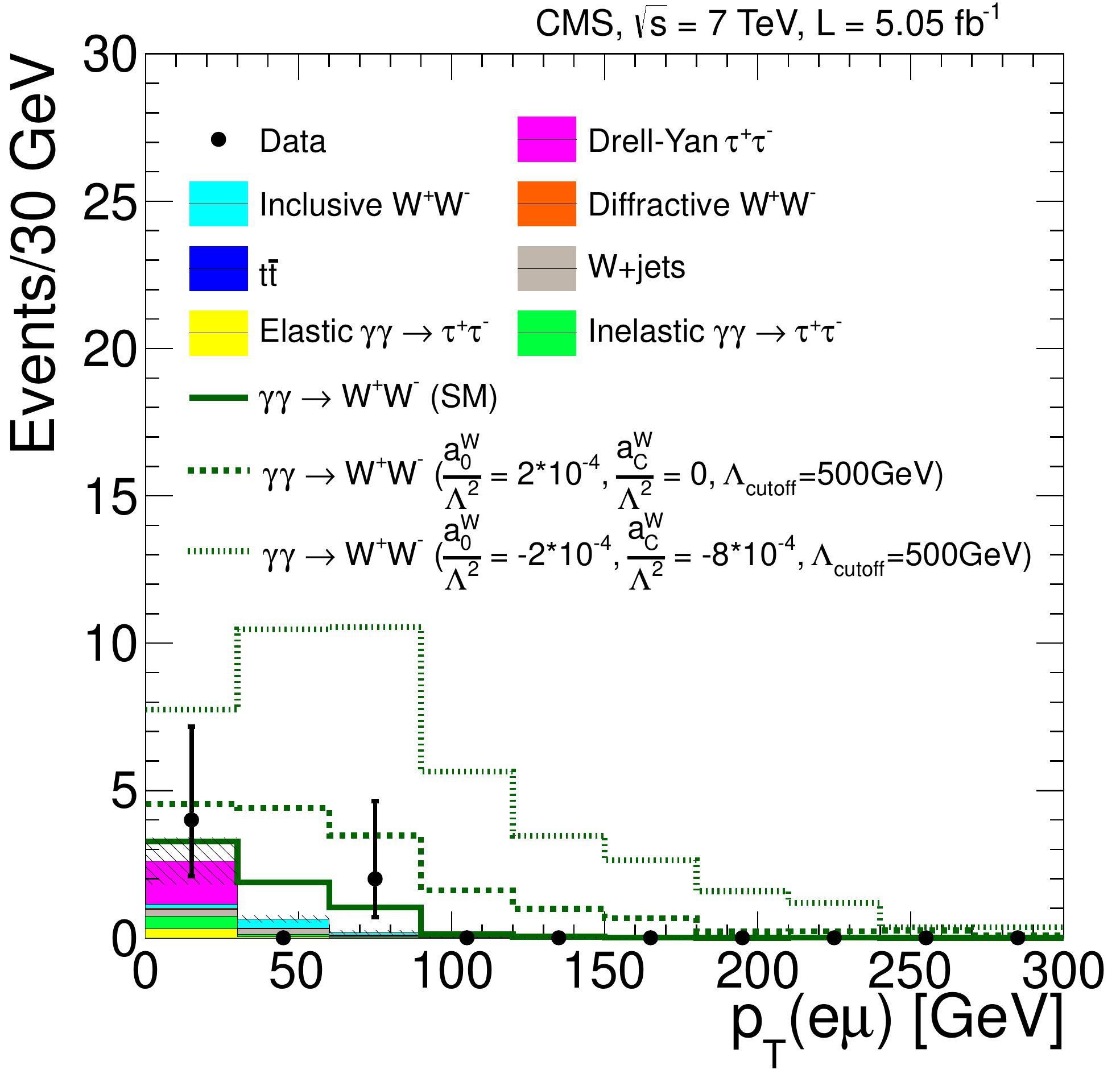}}
    \caption{(a) The extra track multiplicity associated to $\mu^{\pm}e^{\mp}$ vertex and (b) the $p_{{\rm T},\mu^{\pm}e^{\mp}}$ distribution in the exclusive $\gamma\gamma \rightarrow WW$ analysis~\cite{CMS_ggWW}.}
    \label{fig:wwgg}
  \end{center}
\end{figure}

Events are collected using two electron-muon software based trigger algorithms with asymmetric thresholds.
The first algorithm requires a muon with $\pt >$ 17 GeV and an electron with $\pt >$ 8 GeV, while the second requires a muon with $\pt >$ 8 GeV and an electron with $\pt >$ 17 GeV.
In the offline selection reconstructed muons and electrons are required to have opposite charge, $\pt > $ 20 GeV, $|\eta| < $ 2.4 and matched to a primary vertex with zero extra tracks.
The leptons are required to have $m_{\mu^{\pm}e^{\mp}} > $ 20 GeV.
Figure~\ref{fig:wwgg} shows distributions after all the cuts other than the cuts on the variables themselves: (a) extra track multiplicity associated to $\mu^{\pm}e^{\mp}$ vertex with $p_{{\rm T},\mu^{\pm}e^{\mp}} > $ 30 GeV cut, and (b) $p_{{\rm T},\mu^{\pm}e^{\mp}}$.
%In the both figures the cuts on the variable itself is not applied.

A 95\% C.L. upper limit on a fiducial cross-section is derived using the Feldman-Cousins method ~\cite{CMS_ggWW_fc} with the additional cut of $p_{{\rm T},\mu^{\pm}e^{\mp}} > $ 100 GeV as:
\begin{eqnarray}
  \sigma(pp \rightarrow p^{(\ast)}W^{+}W^{-}p^{(\ast)} \rightarrow p^{(\ast)}\mu^{\pm}e^{\mp}p^{(\ast)}) < 1.9 ~ {\rm fb.} \nonumber
\end{eqnarray}
In the same phase space, 95\% C.L. intervals are set following the Feldman-Cousins prescription on the aQGC parameter $a^{W}_{0}$ and $a^{W}_{C}$ ~\cite{CMS_ggWW_aqgc}.
With no form factor and the other parameter fixed to zero they are:
\begin{eqnarray}
  -4.0 \times 10^{-6} < a^{W}_{0}/\Lambda^{2} < 4.0 \times 10^{-6} {\rm ~GeV^{-2}}~(a^{W}_{C}/\Lambda^{2} = 0); \nonumber \\
  -1.5 \times 10^{-5} < a^{W}_{C}/\Lambda^{2} < 1.5 \times 10^{-5} {\rm ~GeV^{-2}}~(a^{W}_{0}/\Lambda^{2} = 0), \nonumber
\end{eqnarray}
where $\Lambda$ is the energy scale of possible new physics.

\newpage
\subsection{$WW\gamma/WZ\gamma$ production}\label{CMS_VVV}

Triboson production offers a direct access to (a)QGCs.
The CMS experiment performed a search for $WW\gamma$ and $WZ\gamma$ production using single lepton final state, which includes $W(\rightarrow \ell \nu) W(\rightarrow jj) \gamma$ and $W(\rightarrow \ell \nu) Z(\rightarrow jj) \gamma$ processes where $\ell = e,\mu$~\cite{CMS_VVV}.
The hadronic decay mode is chosen because of its substantially larger branching ratio than that of the leptonic decay mode.
$WW\gamma$ and $WZ\gamma$ are treated as a single combined signal due to a limitation in the dijet mass resolution.
The amount of data corresponds to $19.3 \pm 0.5~(19.2 \pm 0.5)~{\rm fb}^{-1}$ in the muon (electron) channel, and are taken at $\sqrt{s} = 8$ TeV.

The data are collected with single lepton triggers using $\pt$ thresholds of 24 GeV for muons and 27 GeV for electrons, and the following offline selections are applied:
either one muon with $\pt > 25$ GeV and $|\eta| < $ 2.1 or one electron with $\pt > 30$ GeV and $|\eta| < $ 2.5, excluding the transition region ($1.44 < |\eta| < 1.57$), is required;
events with additional leptons are vetoed;
$\met > 35$ GeV is required as well as requiring the transverse mass of $W$ boson candidate ($\sqrt{\pt^{\ell}\met(1-{\rm cos}(\Delta \phi_{\ell\met}))}$, where $\phi_{\ell\met}$ is the azimuthal angle between the lepton and $\met$ direction) larger than 30 GeV;
at least two jets are reconstructed by anti-$k_{t}$ algorithm with a distance parameter $R =$ 0.5, and satisfy $\pt > $ 30 GeV and $|\eta| < 2.4$;
the highest $\pt$ jet candidates are chosen to form the hadronically decaying boson with $70 < m_{jj} < 100$ GeV requirement applied, and the separation in pseudo-rapidity between the jets of $|\Delta \eta_{jj}| < 1.4$ imposed;
the two jets fail a $b$-tagging requirement with about 70\% efficiency and a rejection factor of 100 to reject top-quark production;
the photon candidate must satisfy $E_{\rm T} >$ 30 GeV and $\eta <$ 1.44;
the $|M_{Z} - m_{e\gamma}| > $ 10 GeV cut is applied to suppress the $Z \rightarrow e e$ background.

\begin{figure}[htp]
  \begin{center}
    \subfigure[]{\includegraphics[width = 0.44\textwidth]{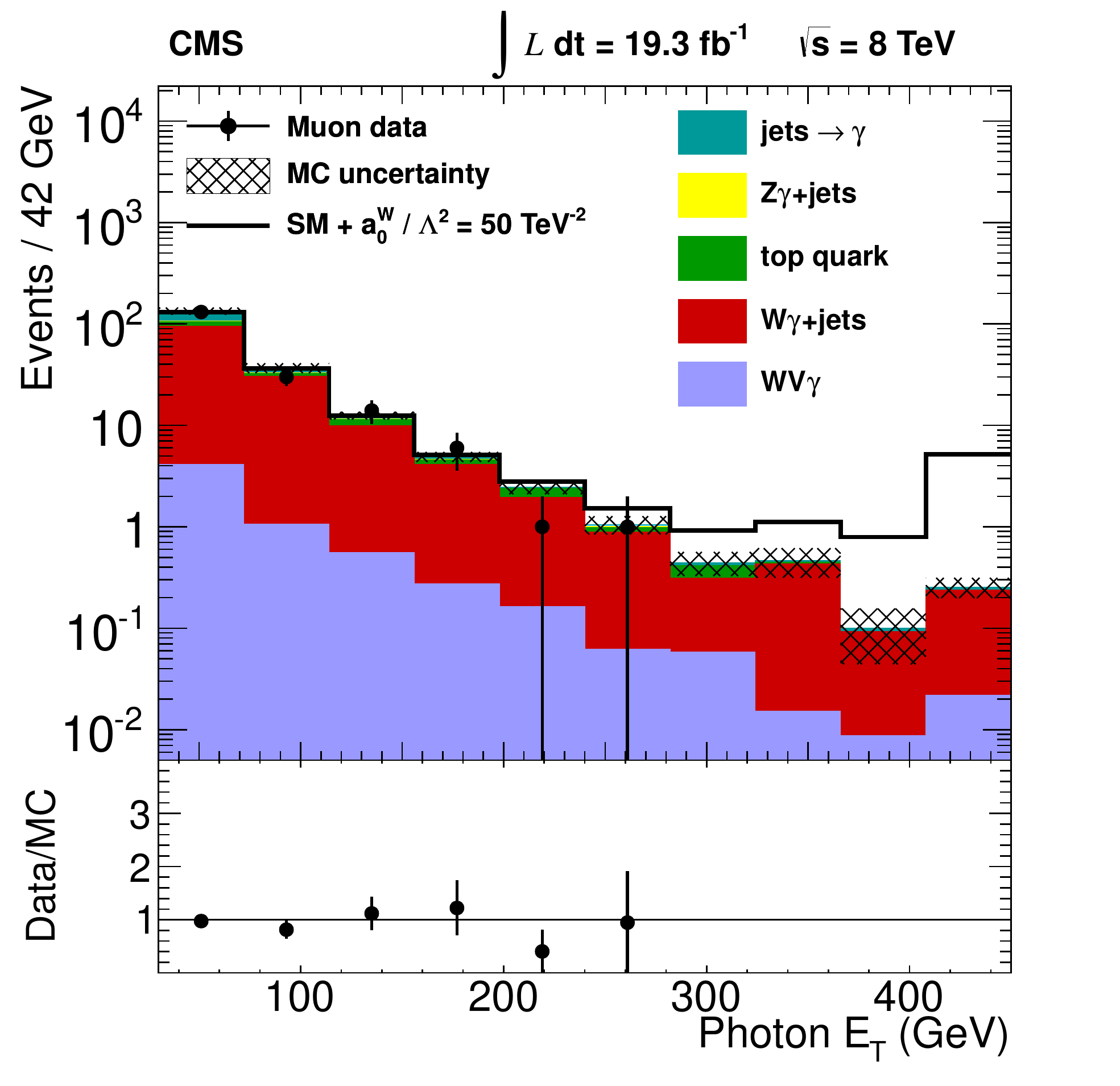}}
    \subfigure[]{\includegraphics[trim=0cm 0cm 4cm 0cm, width = 0.44\textwidth]{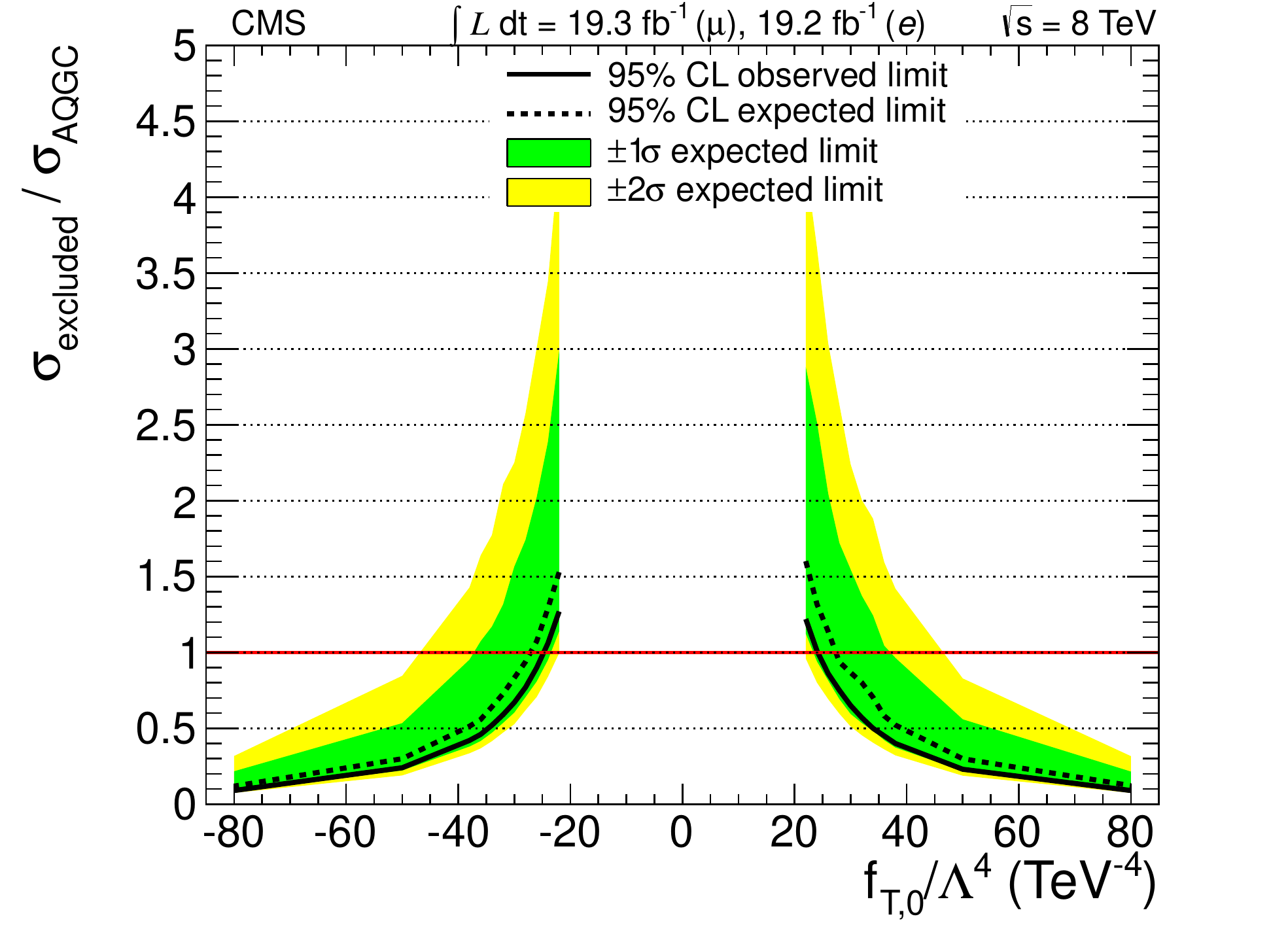}}
    \caption{(a) The photon $E_{T}$ distributions in muon channel and (b) the exclusion limits on $f_{T,0}/\Lambda^{4}$ in the $WW\gamma/WZ\gamma$ analysis~\cite{CMS_VVV}.}
    \label{fig:vvv}
  \end{center}
\end{figure}

A 95\% C.L. observed upper limit of 311 fb is calculated for inclusive cross-section, which is about 3.4 times larger than the SM prediction.
Also, 95\% C.L. upper limits are derived on aQGCs with dimension-6 and dimension-8 parameters~\cite{CMS_VVV_aqgc1,CMS_VVV_aqgc2,CMS_VVV_aqgc3} using photon $E_{\rm T}$ distribution.
%A profile likelihood asymptotic approximation method~\cite{CMS_VVV_lim}, which takes the distributions from the two channels as independent inputs to be combined statistically into a single result, is utilised.
Figure~\ref{fig:vvv} (a) shows the photon $E_{T}$ distributions in muon channel, along with the predicted signal from $WW\gamma\gamma$ aQGC for $a^{W}_{0}/\Lambda^{2}$ = 50 ${\rm TeV}^{-2}$.
%The last bin includes the overflow.
Figure~\ref{fig:vvv} (b) shows the observe and expected exclusion limits for the combination of muon and electron channels for a dimension-8 aQGC parameter $f_{T,0}/\Lambda^{4}$.

%\begin{figure}[htp]
%  \begin{center}
%    \subfigure[]{\includegraphics[width = 0.48\textwidth]{figures/vvv/Table4_aQGClimits.pdf}}
%    \subfigure[]{\includegraphics[width = 0.48\textwidth]{figures/vvv/Table5_FMi_limits.pdf}}
%    \caption{}
%    \label{fig:vvv_tab}
%  \end{center}
%\end{figure}

\begin{figure}[htp]
  \begin{center}
    \subfigure[]{\includegraphics[width = 0.98\textwidth]{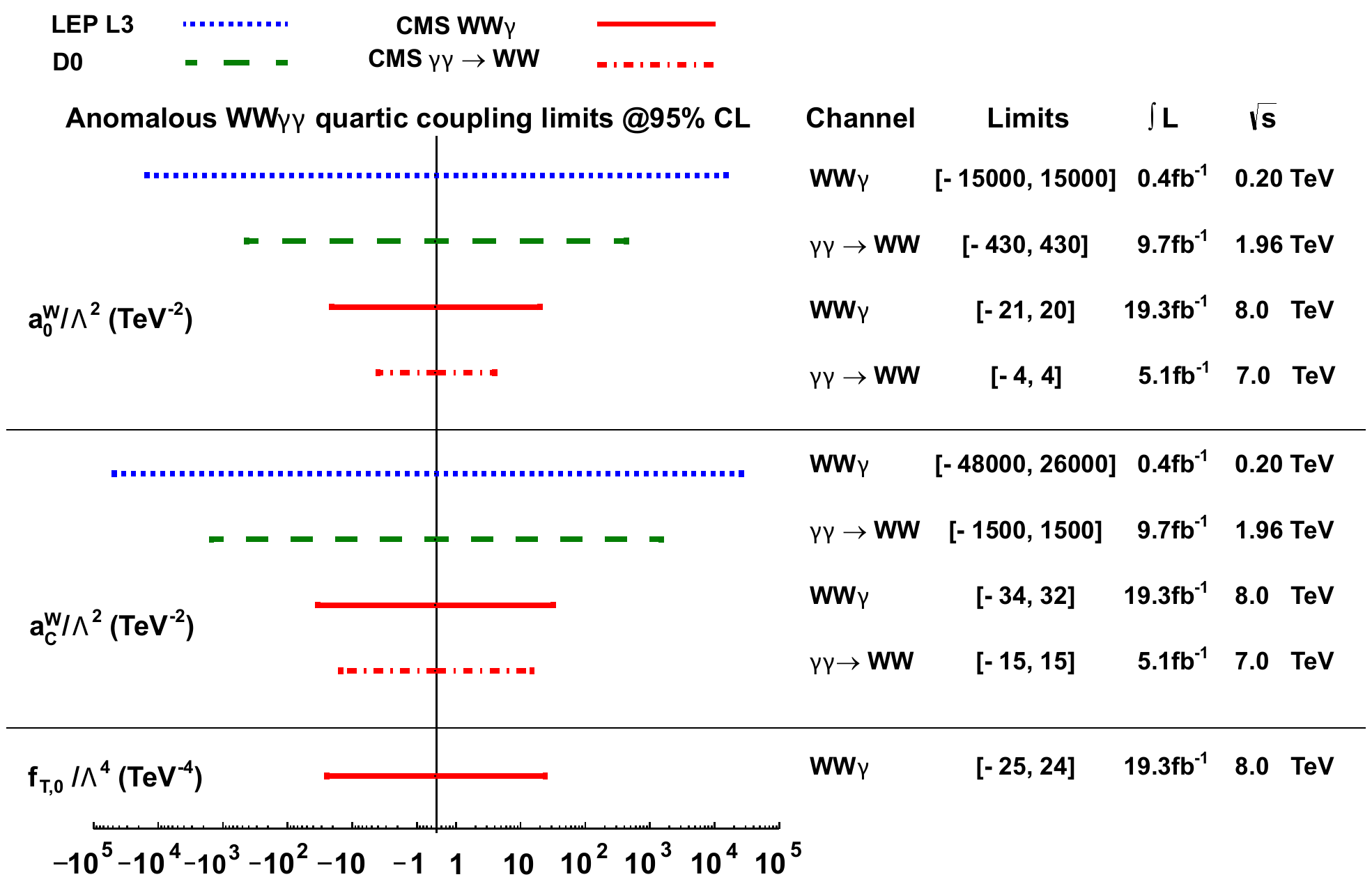}}
    \caption{Comparison of the limits on the $WW\gamma\gamma$ aQGC parameters obtained by the CMS, L3 and D0 collaborations~\cite{CMS_VVV}. The results of the $WW\gamma/WZ\gamma$ analysis is indicated as $WW\gamma$.}
    \label{fig:vvv_res}
  \end{center}
\end{figure}

Figure~\ref{fig:vvv_res} presents exclusion limits on $WW\gamma\gamma$ aQGC parameters obtained by the CMS experiment, including the exclusive $\gamma\gamma \rightarrow WW$ measurement, compared to past measurements performed by L3~\cite{CMS_VVV_l3} and D0~\cite{CMS_VVV_d0}.
The two measurements by the CMS experiment set the world-best exclusion limits.

\subsection{VBF $Z$ boson production with two jets }\label{EWZjets}.

The electroweak $Zjj$ production, which is the production of a $Z$ boson in association with two jets via the $t$-channel exchange of a electroweak gauge boson, includes the contributions from the $Z$ boson production via VBF.
Therefore the process is of particular interest because of the sensitivity to anomalous TGCs (aTGCs) in $WWZ$ vertex.

%The dominant production mechanism for a leptonically decaying $Z/\gamma^{\ast}$ boson in association with two jets ($Zjj$) at the LHC is via the Drell-Yann process, with the additional jets arising as a result of the strong interaction.
%Production of $Zjj$ events via the $t$-channel exchange of an electroweak gauge boson is a purely electroweak process and is therefore much rarer.
%Electroweak $Zjj$ production in the leptonic decay channel is defined to include all contributions to $\ell^{+}\ell{-}jj$ production for which there is a $t$-channel exchange of an electroweak gauge boson~\cite{ATLAS_EWZjets_vbf1,ATLAS_EWZjets_vbf2}.
%These contributions include $Z$ boson production via VBF, which is of particular interest because of the sensitivity to aTGC in $WWZ$ vertex.

\begin{figure}[htp]
  \begin{center}
    \subfigure[]{\includegraphics[width = 0.45\textwidth]{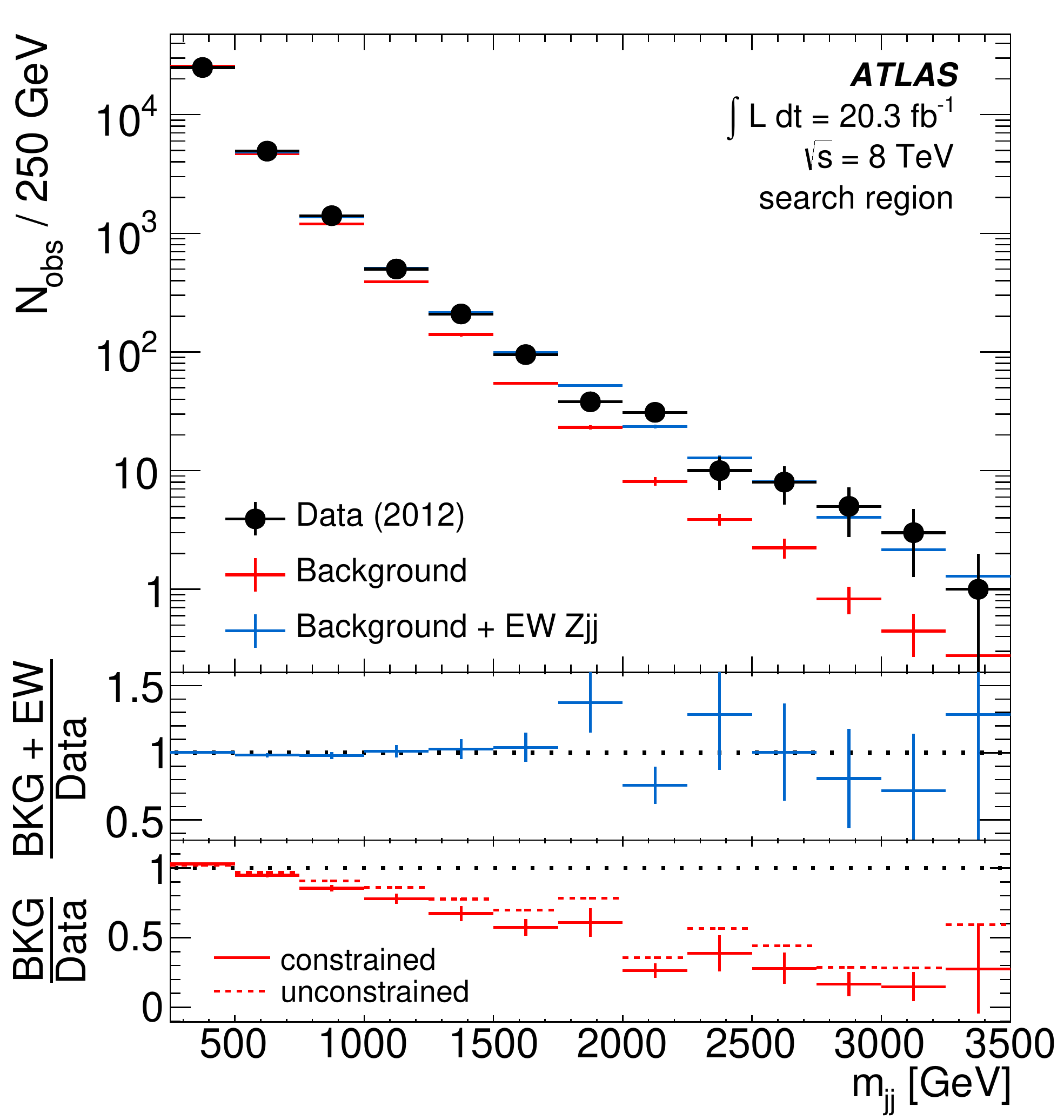}}
    \subfigure[]{\includegraphics[width = 0.48\textwidth]{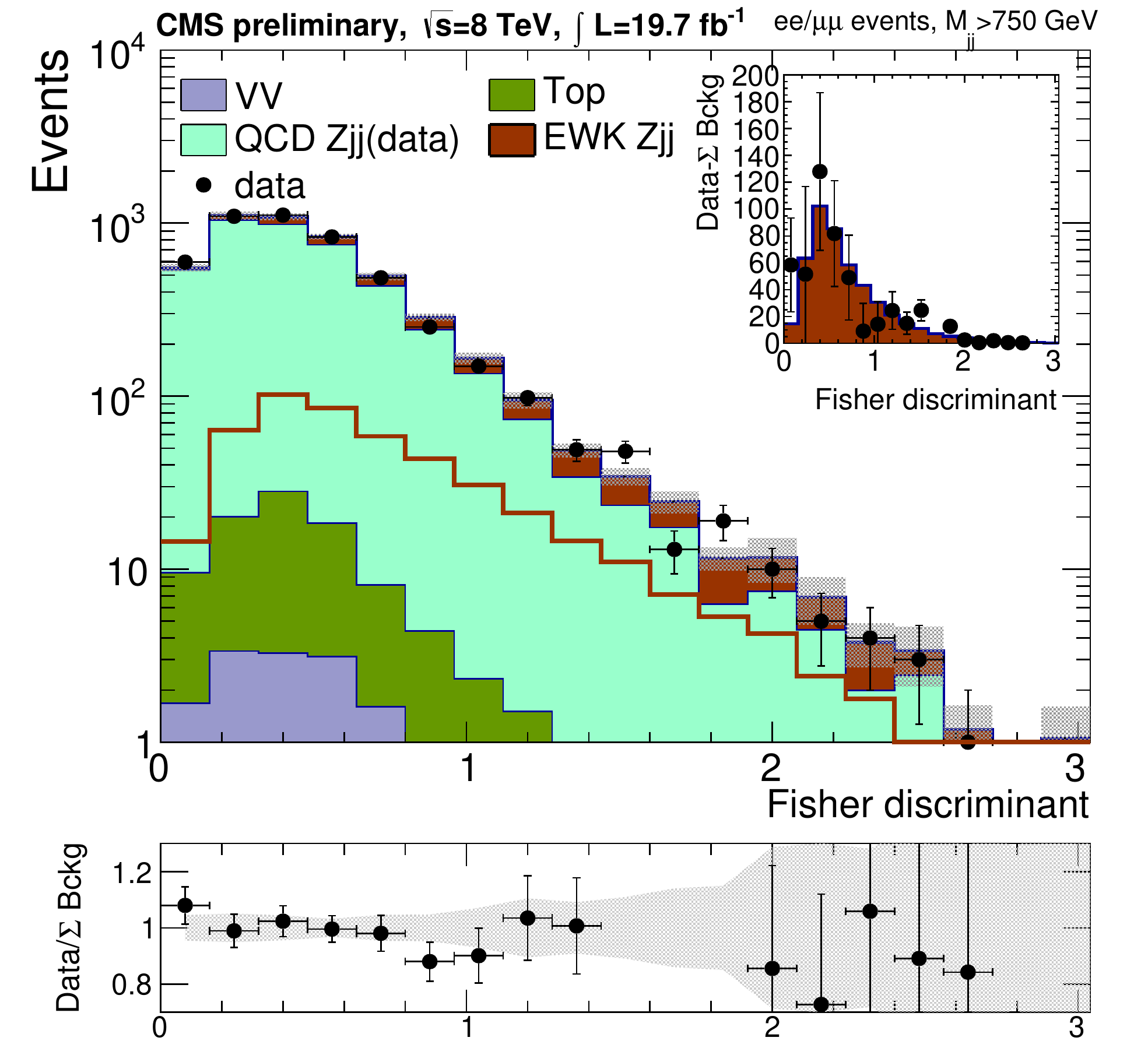}}
    \caption{(a) The $m_{jj}$ distributions in the signal enhanced region by the ATLAS~\cite{ATLAS_EWZjets}. (b) The Fisher discriminant variable obtained with a data driven method by the CMS experiment in the $Zjj$ analyses~\cite{CMS_EWZjets}.}
    \label{fig:ewzjj}
  \end{center}
\end{figure}

The ATLAS and CMS experiments performed measurements of fiducial cross-sections for electroweak $Zjj$ production using 20.3 ${\rm fb}^{-1}$ and  19.7 ${\rm fb}^{-1}$ of proton-proton collision data taken at $\sqrt{s} = 8$ TeV, respectively~\cite{ATLAS_EWZjets,CMS_EWZjets}.
The ATLAS experiment constructed five fiducial regions, and the electroweak component is extracted by a fit to the $m_{jj}$ distribution in a fiducial region chosen to enhance the electroweak contribution over the dominant background of strong $Zjj$ production.
Figure~\ref{fig:ewzjj} (a) shows the $m_{jj}$ distribution in data compared to the unfolded MC predictions in that region.
%The electroweak cross sections measured in two fiducial regions are in good agreement with the SM expectations.
The CMS experiment utilises two independent methods, one is based on a simulation and the other is based on a data driven method in modeling the strong $Zjj$ background.
The results from both measurements are statistically combined.
Figure~\ref{fig:ewzjj} (b) shows the Fisher discriminant variable obtained using the data driven method.
By these measurements, both the ATLAS and CMS experiments independently reject the background-only hypothesis with significance above the 5 $\sigma$ level.

The ATLAS experiment also sets 95\% C.L. limits on the aTGCs on $WWZ$ vertex.
In the standard hadron collider analyses, aTGC limits are set by measuring diboson production, for which all three gauge bosons entering the $WWZ$ vertex have time-like four-momentum.
In the VBF diagram, however, two of the gauge bosons entering the $WWZ$ vertex have space-like four-momentum transfer.
Electroweak $Zjj$ production therefore offers a complementary test of aTGCs, because the effects of boson propagators present in electroweak $Zjj$ production are different from those in diboson production~\cite{ATLAS_EWZjets_vbf3}.
Using the LEP parameterisation~\cite{ATLAS_EWZjets_lep}, the limits are derived as:

\begin{eqnarray}
  -0.50 < \Delta g_{1,Z} < 0.26~({\rm obs.})&,&-0.45 < \Delta g_{1,Z} < 0.22~({\rm exp.}); \nonumber \\
  -0.15 < \lambda_{Z} < 0.13~({\rm obs.})&,&-0.14 < \lambda_{Z} < 0.11~({\rm exp.}). \nonumber
\end{eqnarray}

%Observation of the Higgs Boson,  \cite{Aad:2012tfa},\cite{Chatrchyan:2012ufa}. 

%\begin{figure}[htb]
%\centering
%\includegraphics[height=2in]{head_lhcp2014.jpg}
%\caption{ Place the caption here}
%\label{fig:figure1}
%\end{figure}

%See Figure \ref{fig:figure1} and Table \ref{tab:table1}. 

%\begin{table}[t]
%\begin{center}
%\begin{tabular}{l|ccc}  
%Patient &  Initial level($\mu$g/cc) &  w. Magnet &  
%w. Magnet and Sound \\ \hline
% Guglielmo B.  &   0.12     &     0.10      &     0.001  \\
% Ferrando di N. &  0.15     &     0.11      &  $< 0.0005$ \\ \hline
%\end{tabular}
%\caption{ place the caption here }
%\label{tab:table1}
%\end{center}
%\end{table}

\section{Conclusions}

The investigation of TGCs and QGCs offers an important test of the SM, and is essential in the understanding of electroweak symmetry breaking sector. 
The ATLAS and CMS experiments have provided new exclusion limits using triboson production, VBS, and VBF.
With triboson production and VBS the two LHC experiments have set world-best exclusion limits on aQGCs, while with VBF, a new window in the study of aTGCs with the contributions from space-like gauge bosons has been opened up.
After upgrading the center-of-mass energy and luminosity of the LHC more accurate measurements, that are enough to constrain TeV scale physics, would be delivered~\cite{conclusions}.

%%  if necessary
\Acknowledgements

I would like to thank the conveners of the Standard Model group of the ATLAS and CMS experiments, most notably Dr. Jake Searcy, Dr. Alexander Savin and Dr. Anja Vest for their kind advice on details of the analyses.
I am also grateful to Dr. Alex Read for his help in preparing this conference contribution.

\end{document}